\documentclass[prl,twocolumn,showpacs,preprintnumbers,amsmath,amssymb]{revtex4}
\usepackage{graphicx}
\usepackage{dcolumn}
\usepackage{bm}
\begin{document}

\title{Two Impurity Anderson problem: \\
Perturbative Kondo-Doublet interaction.}
\author{J. Simonin}
\affiliation{Centro At\'{o}mico Bariloche and Instituto Balseiro, \\
8400 S.C. de Bariloche, R\'{i}o Negro,  Argentina}
\date{mayo 2008}

\begin{abstract}
We reobtain the Kondo-Doublet interaction by means of the Brillouin-Wigner Perturbation theory. By applying the same method to the single impurity case we show that the Kondo-Doublet interaction is a direct consequence of the Kondo screening cloud. We fully confirm our previous results on this novel interaction. In particular, we found that for values of the system parameters typical of semiconductor Quantum Dots heterostructures, the ferromagnetic Kondo-Doublet interaction dominates over the RKKY one. 

\end{abstract}
\maketitle
\section{Introduction}

The hybridization between a localized electron and itinerant electrons of a Fermi sea leads to the Kondo effect\cite{hewson}. The itinerant electrons not only screen an impurity spin, leading to the Kondo effect, but also give rise to the Ruderman-Kittel-Kasuya-Yosida (RKKY) interaction between localized spins\cite{kittel}(p.360)\cite{glazm, riera}. The interplay between the Kondo screening and the RKKY interaction remains at the focus of the investigation of strongly correlated electron systems and may play an important role in the heavy fermion metals\cite{coleman}. Moreover, the possibility to use Kondo circuits (made of adatoms\cite{adatom} or quantum dots\cite{sasaki}) as quantum computers\cite{jsqdqw} makes the understanding of the interaction between ``magnetic impurities" a priority.

The competition between Kondo screening and the RKKY is depicted by the well-known Doniach's phase diagram\cite{doni}. It is drawing by comparing the single-impurity Kondo energy, $ \delta_K \propto exp[-1 / 2 \rho_o J ] $ , with the RKKY interaction ($ \propto \rho_o J^2$). Here, $\rho_o$ is the density of states of the conduction electrons at the Fermi level ($E_F$), and $J$ is the effective Kondo coupling. Based on the single-impurity Kondo results\cite{hewson}, it is generally believed that the Kondo effect screens the impurities spin whereas that the RKKY give rise to magnetic structures. One thing does not fit in the above picture: the extent ($\xi_K$) of the Kondo screening cloud. This length ($\simeq \lambda_F E_F/\delta_K$) is much larger than the Fermi wavelength, $\lambda_F$, which is the typical range of the RKKY interaction. Thus, cooperative Kondo effects should appear well in advance of the RKKY effects \cite{coleman,coleman2}. Failure to discern those effects has lead to question the existence of the Kondo cloud\cite{coleman2}.   

The Kondo-Doublet interaction\cite{jsfull} fills the holes in the picture above. It shows that a cooperative Kondo effect effectively takes place between pairs of magnetic impurities, and that, while screening the total spin, it generates a strong ferromagnetic correlation between the localized spins. 

Here we reobtain the Kondo-Doublet (KD) interaction by means of the Brillouin-Wigner (BW) perturbative method\cite{ziman}(p.53), and, by comparing the electron process in it with that in the single impurity screening cloud, we show that the KD interaction is a direct consequence of the Kondo screening cloud. We also introduce the updated 1-dim Doniach phase diagram, which is relevant for technological applications\cite{jsqdqw}, and we discuss briefly the conceptual differences between BW and Rayleigh-Schr\"{o}dinger perturbation theories.

\section{Hamiltonian and BW Theory}

The Anderson Hamiltonian is given by the sum of the metal host, magnetic impurities, and hybridization terms, 
\begin{eqnarray}\label{ho}
H= \sum_{k \sigma} e_k c^\dag_{k\sigma} c_{k\sigma}- E_d \sum_{j \sigma} d^\dag_{j\sigma} d_{j\sigma}  + U \sum_j n_{j\downarrow}n_{j\uparrow} \nonumber \\
 + \ \textbf{v}\sum_{j\ k\sigma} (e^{i \textbf{k}. \textbf{R}_j} 
d^\dag_{j\sigma} c_{k\sigma}+\ h.c. \ ) \ \ \ \ \ \ \ \ ,
\end{eqnarray}
where the fermion operators $c_{k\sigma }\ (d_{j\sigma })$ correspond to the  metal band states  (the impurity at $\textbf{R}_j$). Single state energies $e_k, -E_d$ are measured from the Fermi level, and $\textbf{v}= V/\sqrt{N}$ is the \textit{c-d} hybridization divided by the square root of the number of band states. In the Kondo limit the impurity level is well below the Fermi energy and the Coulomb repulsion $U$ at the impurity is strong. The hybridization makes each impurity to resonate between its possible populations,  $ d^2\leftrightarrows d^1 \leftrightarrows d^0$, while generating excitations in the band. We analyze in the following the $E_{d^2} \gg  E_{d^0} \gg E_{d^1}$ case, i.e. $-2E_d+U \gg 0 \gg -E_d$, therefore we retain only the $d^1 \leftrightarrows d^0$ hybridization channel. The Fermi sea $|F\rangle$ is taken as the vacuum and we make an electron-hole transformation for band states below the Fermi level, $b^\dag_{k\sigma}\equiv c_{k\overline{\sigma}}$ for $|k|\leq k_F$. Note that the energy of a hole excitation is explicitly positive.

We consider here the Two Impurity Anderson (TIA) case, one impurity placed at $-R/2$ and the other at $R/2$, over the $x$-axis. We use a ``ket" notation for the impurity configurations: the first symbol in the ket indicates the left impurity status (the one at $x=-R/2$)  and the second one the status of the impurity on the right, e.g. $\ | 0 ,\uparrow \rangle \equiv d^\dag_{R\uparrow}|F\rangle$, $|\!\! \downarrow , \uparrow \rangle \equiv d^\dag_{L\downarrow} d^\dag_{R\uparrow} |F\rangle$.

To analyze the system we use Brillouin-Wigner perturbation theory (Ziman\cite{ziman},p.53). At first order, for a given initial state $|\psi_0 \rangle$, of energy $E_0$, BW theory gives the following expressions for the corrected energy and wave function,
\begin{eqnarray}
E & = & E_0 + \sum_a \frac {V_{0a}V_{a0}}{E-E_a}  \ \ \label{ebwa} ,   \\
|\psi \rangle & = & |\psi_0 \rangle + \sum_a \frac {V_{a0}}{E-E_a} \ |\psi_a \rangle \ \ \label{ebwb},
\end{eqnarray}
where $|\psi_a \rangle$ are the states obtained \textit{via} the application of the non-diagonal terms of the Hamiltonian to the initial state, $E_a$ their energy, and $V_{a0} = \langle \psi_a| H_V |\psi_0 \rangle$ the corresponding matrix element.

\section{Kondo-Doublet interaction}

To obtain the Kondo-Doublet interaction we apply BW perturbation theory to the $S=1/2$ subspace of the TIA system. For the spin-up odd subspace we take
\begin{equation}\label{fi0}
   |\psi_0 \rangle = \frac {d^\dag_{R\uparrow} - d^\dag_{L\uparrow}}{\sqrt{2}} \ |F\rangle =\frac {|0,\uparrow\rangle - |\!\uparrow, 0\rangle}{\sqrt{2}}  \ \ ,
\end{equation}
of energy $E_0=-E_d$, and the following $\psi_a$'s
\begin{equation}\label{fia}
  b^\dag_{k \uparrow} \ |\!\uparrow, \downarrow \rangle \ \ \ \ , 
 \ \ b^\dag_{k \uparrow} \ |\!\downarrow, \uparrow \rangle \ \ \ \ , 
 \ \ b^\dag_{k \downarrow} \ |\!\uparrow, \uparrow \rangle \ \ \ \ ,
\end{equation}
all of them of energy $E_a=-2E_d+e_k$. They correspond to the promotion of one electron from below $E_F$ , \textit{via} the action of the hybridization term, to the impurity that is empty in $|\psi_0 \rangle$. Note that the $b^\dag_{k \downarrow} |\!\uparrow, \uparrow \rangle$ configuration is obtained from either of the components of the $|\psi_0 \rangle$ state, we will see that the ensuing interference generates the Kondo-Doublet interaction. Their $V_{a0}$ elements are given by  
\begin{equation}\label{v0a}
 \ \ e^{-i \textbf{k.R}/2} \ \ , \ \ e^{+i \textbf{k.R}/2} \ \ ,  \ \ (e^{-i \textbf{k.R}/2}\ + e^{+i \textbf{k.R}/2})  \ \ , 
\end{equation}
times $\textbf{v}/\sqrt{2}$, respectively. Application of BW Eq.(\ref{ebwa}) thus gives
\begin{equation}\label{ed1}
E=-E_d + \textbf{v}^2 \sum_k \frac{ 1 + 2 \cos^2{(\textbf{k.R}/2)} }{E-(-2E_d+e_k)} \ \ ,
\end{equation}
where the sum is over the hole states ($|k|\le k_F$). Assuming $E=-2E_d-\delta_o(R)$, and using $2 \cos^2{(x/2)}= 1 + \cos{x}$, Eq.(\ref{ed1}) transforms to 
\begin{equation}\label{ed2}
    E_d + \delta_o(R)=  \frac{V^2}{N} \sum_k \frac{2 + \cos{(\textbf{k.R})} }{\delta_o(R) + e_k} \ \ .
\end{equation}
From this equation, with the usual half-filled flat-band assumptions, the correlation energy of the odd Kondo-Doublet is obtained
\begin{equation}\label{do}
\delta_{o(e)}(R) = D\ \exp{(-1/(2\pm C_h(R)) J_n)} \ \ ,
\end{equation}
where $J_n = \rho_0 V^2/E_d $ is the effective Kondo coupling times the density of band states, $D$ is the half-band width and $C_h(R)$ is the \textit{hole}-coherence factor, which is given by
\begin{equation}\label{ch0}
C_h(R)=  \sum_k \frac{\cos{(\bf{k.R})}}{\delta_o(R)+e_k}\ /\  \sum_k \frac{1}{\delta_o(R)+e_k} \ \ ,
\end{equation}
and it is, as a function of $R$, an oscillating decaying function ( $|C_h(R)|\leq1$, of period $\simeq \lambda_F$, and $C_h(0)=1$, $C_h(\infty)=0$)\cite{jsfull,jscloud}. Its range determines the range of the Kondo-doublet interaction, \textit{i.e.} how close the impurities must be in order to significatively interact through this mechanism. See that for $R=0$ we have $\delta_{o}(0) = D\ \exp{(-1/(3 J_n))} \ \gg \delta_K $. The minus sign in Eq.(\ref{do}) corresponds to the even Kondo-Doublet correlation energy $\delta_e$, which is the dominant one for the regions in which $C_h(R)<0$. 

BW Eq.(\ref{ebwb}) for the wave function gives   
\begin{eqnarray}\label{wodd0}
|D_{o\uparrow}\rangle= \frac{|0,\uparrow\rangle - |\!\uparrow, 0\rangle}{\sqrt{2}} \  + \ \frac{\textbf{v}}{\sqrt{2}} \ \sum_{k}\ \frac{1}{\delta_o(R) +e_k} \nonumber \\  ( \ e^{-i \textbf{k.R}/2} \  b^\dag_{k \uparrow} \ |\!\uparrow, \downarrow \rangle \ + \ e^{+i \textbf{k.R}/2} \ b^\dag_{k \uparrow}  \ |\!\downarrow, \uparrow \rangle  \nonumber \\ + \ ( \ e^{-i \textbf{k.R}/2}\ + \ e^{+i \textbf{k.R}/2} \ ) \ b^\dag_{k \downarrow}  \ |\!\uparrow, \uparrow \rangle \ )  \ \ \ , 
\end{eqnarray}
\textit{i.e.} the result of the variational ansatz used in Ref.\cite{jsfull}. Thus Brillouin-Wigner perturbation theory fully confirms our previous findings on the ferromagnetic Kondo-Doublet interaction.   

To unveil the physics behind the Kondo-Doublet interaction we apply now BW theory to the single impurity Anderson model\cite{hewson}. If only the impurity at $R_L=- R/2$ is present, one takes $\psi_0 = |F\rangle$ and $\psi_a's$ equal to $\ b^\dag_{k\uparrow} \ |\!\downarrow, - \rangle$, and $\ b^\dag_{k\downarrow } \ |\!\uparrow, - \rangle$, were the $-$ symbol at the right in the ket indicates that the impurity at $+R/2$ is absent. Their $V_{a0}$ elements are all equal to $\textbf{v} \ e^{+i \textbf{k.R}/2} $. Thus, application of BW theory to this singlet gives 
\begin{equation}\label{sek1}
    E_S= \frac{V^2}{N} \sum_k \frac{2}{E_S -(-E_d+e_k)} \ ,
\end{equation}
which upon assuming $E_S=- E_d-\delta_K$, gives the Kondo correlation energy $\delta_K = D\ \exp{(-1/(2 J_n))}$. For the singlet wave function one obtains  
\begin{equation}\label{wsk}
|S_L\rangle= |F\rangle +  \sum_{k} \ Z_k \ (  b^\dag_{k \uparrow} |\!\downarrow,-\rangle +  b^\dag_{k \downarrow} |\!\uparrow,- \rangle )  \ \ , 
\end{equation}
were $Z_k = \textbf{v} \ e^{+i \textbf{k.R}/2}/(\delta_K+e_k)$, \textit{i.e.} the well-known Varmat-Yafet (VY) Kondo singlet\cite{varma,gunna} , but for an impurity at $-R/2$ instead of at the origin. Comparing the VY singlet with our KD ansatz is easy to see that the latter can be written as 
\begin{equation}\label{wds}
 \ |D_{o\uparrow}\rangle = \frac{|S_L\rangle \otimes |-,\!\uparrow \rangle - |S_R\rangle \otimes |\!\uparrow,\!-\rangle} {\sqrt{2}} \ \  , 
\end{equation}
taken that $ |\sigma,-\rangle \otimes |-,\sigma'\rangle = |\sigma,\sigma'\rangle $ and $ |F\rangle \otimes |-,\sigma'\rangle = |0,\sigma'\rangle $. 

In fact, for each single impurity one can define Kondo hole orbitals\cite{jscloud,berg1} localized around the impurity,   
\begin{equation}\label{kjorb}
 K^\dag_{j\sigma} = \frac{\textbf{v}}{\sqrt{n_\delta}} \sum_k \ \frac{e^{-i \textbf{k.R}_j}} {\delta + e_k} \ \ b^\dag_{k\sigma}, 
\end{equation}
where $n_\delta \simeq J/2\delta \gg 1 $. The corresponding VY singlets are thus given by $|S_j\rangle = |F\rangle + \sqrt{n_\delta}\ (K^\dag_{j\uparrow} d^\dag_{j\downarrow}+K^\dag_{j\downarrow} d^\dag_{j\uparrow})\ |F\rangle$. It cost kinetic energy to form these orbitals. All the energy gain, the Kondo energy plus the cost of the orbitals, comes from the resonance between the orbitals, which use the Fermi sea configuration as a nearly virtual bridge. This resonance of a hole between the spin-up and spin-down Kondo orbitals screens the impurity spin. Localization of these orbitals ($\sim \xi_K$) is ruled by the $\delta$ ($= \delta_K$ or $\delta_o, \delta_e$) term in the denominator of the amplitudes, Eq.(\ref{kjorb}). The \textit{hole}-coherence factor $C_h(R)$ of the KD interaction, Eq.(\ref{ch0}), is proportional to the amplitude of the Kondo orbital of one impurity evaluated at the other one, thus confirming that the KD interaction is produced directly by the resonance of the Kondo cloud at both impurities, Eq.(\ref{wds}).

BW theory can also be applied to the $|\!\!\uparrow , \downarrow \rangle \pm |\!\!\downarrow , \uparrow \rangle$ states\cite{jsfull}. Following it to second order in the wave function, one obtains the RKKY result: the plus combination gains an energy $\Sigma(R)$, and the minus combination losses the same amount. They are the $S_z=0$ component of the ferromagnetic triplet, and the $S=0$ antiferromagnetic singlet, respectively. Twice $\Sigma(R)$ is the RKKY interaction energy. Therefore, for these states BW result coincide with that of standard Rayleigh-Schr\"{o}dinger (RS) power expansion perturbation theory. But with an important difference: in the RS scheme the weight in the wave function of the intermediate configurations $b^\dag_{k \downarrow} c^\dag_{q \downarrow} |\!\!\uparrow , \uparrow \rangle$ diverges due to the $(E_0-E_2)=-(e_k+e_q)$ denominator that appears in them, whereas for the BW theory that denominator remains finite because it changes to $(E-E_2)$. The RKKY interaction is generated by the exchange of $b^\dag_{k \sigma} c^\dag_{q \sigma}$ pairs between the impurities, \textit{i.e.} the interference between the effects of the Fridel oscillations\cite{hewson,berg3,affle2} generated by one of them in the other one.

One of the main topics in magnetic impurities is the RKKY-Kondo interplay, as depicted in Doniach's phase diagram\cite{doni,coleman2,doni2}. Such phase diagram was traditionally drawn by comparing the two impurity RKKY interaction with the single-impurity Kondo energy. But the Kondo-Doublet interaction analysis shows that Kondo-like structures involving two impurities have a stronger correlation energy than the single impurity case. Therefore, a proper Doniach's phase diagram must drawn by comparing the RKKY, which is mediated by the Fridel oscillations, with the KD interaction, which is directly mediated by the Kondo cloud. 

\begin{figure}[h]
\includegraphics[width=\columnwidth]{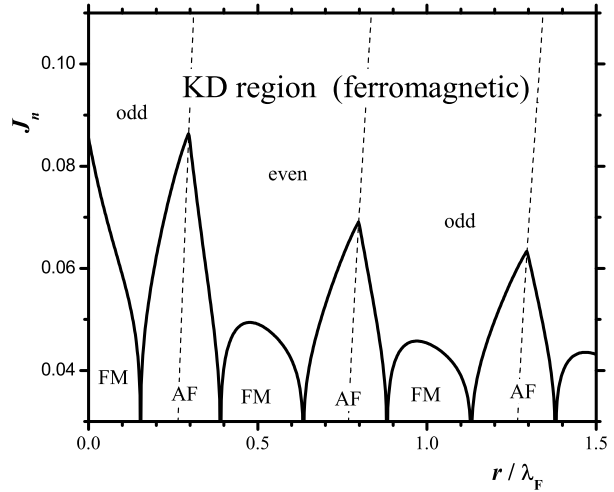}
\caption{\textit{dim}-1 Doniach's phase diagram. The Kondo-Doublet interaction is stronger that the RKKY one in the upper region of the parameter space, which corresponds to values of $J_n$ typical of semiconductors Quantum Dots. The KD  is always ferromagnetic (in both odd and even phases), whereas that the RKKY alternates between odd-ferromagnetic (FM) and even-antiferromagnetic (AF) phases. } \label{fig1}
\end{figure}

In Fig.\ref{fig1} we plot such phase diagram for the  dimension $1$ case, which is relevant for many technological applications\cite{jsqdqw}. It can be seen in Fig.\ref{fig1} that for $J_n \geq 0.085$, which corresponds to $\delta_K \simeq 0.0028 \ D$, the KD is always stronger than the RKKY. The main differences between both interactions are apparent in the figure: the KD ``oscillates" with a period $k_F R$, instead of $2 k_F R$, and , as a function of the distance between the impurities, it decays slowly than the RKKY. Precisely this behavior has been  reported in a recent numerical study of two Anderson impurities embedded in a linear chain\cite{riera}. Together with a quickly decaying $2 k_F R$ oscillation of the correlations, due to the RKKY (as quoted by the authors\cite{riera}) and the superexchange (not quoted), they also found a slowly decaying  $k_F R$ modulation, as predicted by the KD interaction.

\section{Concluding Remarks}

Summarizing, we reobtain the Kondo Doublet interaction using Brillouin-Wigner perturbation theory, fully confirming our previous results. By applying the same method to the single impurity case we show that the KD interaction results from the resonance of the Kondo cloud of one impurity at the other one. This is a clearly distinct mechanism than that of the RKKY interaction, which depends on the Fridel oscillations, \textit{i.e.} the interchange of an \textit{e-h} pair between the impurities.

At this point is worth to point out the differences between Brillouin-Wigner and Rayleigh-Schr\"{o}dinger perturbation theories. RS theory relays in the assumption that the analyzed quantities can be expanded as a power series of a given parameter. That approach works fine with the RKKY interaction, but fails when dealing with Kondo-like structures, which can not be power series expanded\cite{kittel}(p.155)\cite{pwa}. BW theory, instead, makes no previous assumption on the system. BW theory, o more precisely the refined BW theory by Goldhammer and Feenberg\cite{bw2,lipp}, can be seen as an expansion in the number of degree of freedom allowed to the system. But no restriction is applied about how the system uses that degree of freedom\cite{ziman}(p.56)\cite{bethe}. 

Just one degree of freedom is enough to generate the KD interaction: the possibility of a hole to resonate between the impurities (Eqs.(\ref{fi0}-\ref{wodd0})). This is the very same degree of freedom that screens the impurity spin in the single impurity case (Eqs.(\ref{sek1}-\ref{wsk})). Note also that this interaction can be easily generalized to the many impurity case, with the band excitation resonating between all the impurities (an extended Kondo state), making thus a direct connection with the heavy fermions problem.   
 
Therefore, the main result of this study is that the Kondo cloud exits \cite{coleman2,jscloud,berg1}, and that it has a very strong physical consequence: the Kondo-Doublet interaction.

\begin{acknowledgments}

I acknowledge useful talks with G.Chiappe as well as the hospitality of the University of Alicante (Espa\~{n}a), where this study started. I thanks the CONICET (Argentina) for partial financial support.

\end{acknowledgments}

\bibliography{BW}

\begin{thebibliography}{23}
\expandafter\ifx\csname natexlab\endcsname\relax\def\natexlab#1{#1}\fi
\expandafter\ifx\csname bibnamefont\endcsname\relax
  \def\bibnamefont#1{#1}\fi
\expandafter\ifx\csname bibfnamefont\endcsname\relax
  \def\bibfnamefont#1{#1}\fi
\expandafter\ifx\csname citenamefont\endcsname\relax
  \def\citenamefont#1{#1}\fi
\expandafter\ifx\csname url\endcsname\relax
  \def\url#1{\texttt{#1}}\fi
\expandafter\ifx\csname urlprefix\endcsname\relax\def\urlprefix{URL }\fi
\providecommand{\bibinfo}[2]{#2}
\providecommand{\eprint}[2][]{\url{#2}}

\bibitem[{\citenamefont{Hewson}(1993)}]{hewson}
\bibinfo{author}{\bibfnamefont{A.~C.} \bibnamefont{Hewson}},
  \emph{\bibinfo{title}{The Kondo Problem to Heavy Fermions}}
  (\bibinfo{publisher}{Cambridge University Press, Cambridge},
  \bibinfo{year}{1993}).

\bibitem[{\citenamefont{Kittel}(1987)}]{kittel}
\bibinfo{author}{\bibfnamefont{C.}~\bibnamefont{Kittel}},
  \emph{\bibinfo{title}{Quantum Theory of Solids, S.R.P.}}
  (\bibinfo{publisher}{John Wiley Sons, New York}, \bibinfo{year}{1987}).

\bibitem[{\citenamefont{Vavilov and Glazman}(2005)}]{glazm}
\bibinfo{author}{\bibfnamefont{M.~G.} \bibnamefont{Vavilov}} \bibnamefont{and}
  \bibinfo{author}{\bibfnamefont{L.~I.} \bibnamefont{Glazman}},
  \bibinfo{journal}{Phys. Rev. Lett.} \textbf{\bibinfo{volume}{94}},
  \bibinfo{pages}{086805} (\bibinfo{year}{2005}).

\bibitem[{\citenamefont{Costamagna and Riera}(2008)}]{riera}
\bibinfo{author}{\bibfnamefont{S.}~\bibnamefont{Costamagna}} \bibnamefont{and}
  \bibinfo{author}{\bibfnamefont{J.~A.} \bibnamefont{Riera}},
  \textbf{\bibinfo{volume}{08031062}} (\bibinfo{year}{2008}), \bibinfo{note}{to
  appear PRB}.

\bibitem[{\citenamefont{Coleman}(2006)}]{coleman}
\bibinfo{author}{\bibfnamefont{P.}~\bibnamefont{Coleman}},
  \emph{\bibinfo{title}{Heavy Fermions: electrons at the edge of magnetism}}
  (\bibinfo{publisher}{cond-mat/0612006}, \bibinfo{year}{2006}).

\bibitem[{\citenamefont{Knorr et~al.}(2002)\citenamefont{Knorr, Schneider,
  Diekhoner, Wahl, and Kern}}]{adatom}
\bibinfo{author}{\bibfnamefont{N.}~\bibnamefont{Knorr}},
  \bibinfo{author}{\bibfnamefont{M.~A.} \bibnamefont{Schneider}},
  \bibinfo{author}{\bibfnamefont{L.}~\bibnamefont{Diekhoner}},
  \bibinfo{author}{\bibfnamefont{P.}~\bibnamefont{Wahl}}, \bibnamefont{and}
  \bibinfo{author}{\bibfnamefont{K.}~\bibnamefont{Kern}},
  \bibinfo{journal}{Phys. Rev. Lett.} \textbf{\bibinfo{volume}{88}},
  \bibinfo{pages}{096804} (\bibinfo{year}{2002}).

\bibitem[{\citenamefont{Sasaki et~al.}(2006)\citenamefont{Sasaki, Kang,
  Kitagawa, Yamaguchi, Miyashita, Maruyama, Tamura, Akazaki, Hirayama, and
  Takayanagi}}]{sasaki}
\bibinfo{author}{\bibfnamefont{S.}~\bibnamefont{Sasaki}},
  \bibinfo{author}{\bibfnamefont{S.}~\bibnamefont{Kang}},
  \bibinfo{author}{\bibfnamefont{K.}~\bibnamefont{Kitagawa}},
  \bibinfo{author}{\bibfnamefont{M.}~\bibnamefont{Yamaguchi}},
  \bibinfo{author}{\bibfnamefont{S.}~\bibnamefont{Miyashita}},
  \bibinfo{author}{\bibfnamefont{T.}~\bibnamefont{Maruyama}},
  \bibinfo{author}{\bibfnamefont{H.}~\bibnamefont{Tamura}},
  \bibinfo{author}{\bibfnamefont{T.}~\bibnamefont{Akazaki}},
  \bibinfo{author}{\bibfnamefont{Y.}~\bibnamefont{Hirayama}}, \bibnamefont{and}
  \bibinfo{author}{\bibfnamefont{H.}~\bibnamefont{Takayanagi}},
  \bibinfo{journal}{Phys. Rev. B} \textbf{\bibinfo{volume}{73}},
  \bibinfo{pages}{161303} (\bibinfo{year}{2006}).

\bibitem[{\citenamefont{Simonin}(2006{\natexlab{a}})}]{jsqdqw}
\bibinfo{author}{\bibfnamefont{J.}~\bibnamefont{Simonin}},
  \bibinfo{journal}{Phys. Rev. Lett.} \textbf{\bibinfo{volume}{97}},
  \bibinfo{pages}{266804} (\bibinfo{year}{2006}{\natexlab{a}}).

\bibitem[{\citenamefont{Doniach}(1977)}]{doni}
\bibinfo{author}{\bibfnamefont{S.}~\bibnamefont{Doniach}},
  \bibinfo{journal}{Physica B C} \textbf{\bibinfo{volume}{91}},
  \bibinfo{pages}{231} (\bibinfo{year}{1977}).

\bibitem[{\citenamefont{Coleman}(2002)}]{coleman2}
\bibinfo{author}{\bibfnamefont{P.}~\bibnamefont{Coleman}},
  \emph{\bibinfo{title}{Lectures on the Physics of Highly Correlated Electron
  Systems VI}} (\bibinfo{publisher}{Ed. F. Mancini, Am. Inst. Phys., New York},
  \bibinfo{year}{2002}).

\bibitem[{\citenamefont{Simonin}(2006{\natexlab{b}})}]{jsfull}
\bibinfo{author}{\bibfnamefont{J.}~\bibnamefont{Simonin}},
  \bibinfo{journal}{Phys. Rev. B} \textbf{\bibinfo{volume}{73}},
  \bibinfo{pages}{155102} (\bibinfo{year}{2006}{\natexlab{b}}).

\bibitem[{\citenamefont{Ziman}(1969)}]{ziman}
\bibinfo{author}{\bibfnamefont{J.~M.} \bibnamefont{Ziman}},
  \emph{\bibinfo{title}{Elements of Avanced Quantum Theory}}
  (\bibinfo{publisher}{Cambridge University Press, Cambridge},
  \bibinfo{year}{1969}).

\bibitem[{\citenamefont{Simonin}(2007)}]{jscloud}
\bibinfo{author}{\bibfnamefont{J.}~\bibnamefont{Simonin}},
  \bibinfo{journal}{cond-mat} \textbf{\bibinfo{volume}{07083604}}
  (\bibinfo{year}{2007}).

\bibitem[{\citenamefont{Varma and Yafet}(1976)}]{varma}
\bibinfo{author}{\bibfnamefont{C.~M.} \bibnamefont{Varma}} \bibnamefont{and}
  \bibinfo{author}{\bibfnamefont{Y.}~\bibnamefont{Yafet}},
  \bibinfo{journal}{Phys. Rev. B} \textbf{\bibinfo{volume}{13}},
  \bibinfo{pages}{2950} (\bibinfo{year}{1976}).

\bibitem[{\citenamefont{Gunnarsson and Schonhammer}(1983)}]{gunna}
\bibinfo{author}{\bibfnamefont{O.}~\bibnamefont{Gunnarsson}} \bibnamefont{and}
  \bibinfo{author}{\bibfnamefont{K.}~\bibnamefont{Schonhammer}},
  \bibinfo{journal}{Phys. Rev. Lett.} \textbf{\bibinfo{volume}{50}},
  \bibinfo{pages}{604} (\bibinfo{year}{1983}).

\bibitem[{\citenamefont{Bergmann}(2008{\natexlab{a}})}]{berg1}
\bibinfo{author}{\bibfnamefont{G.}~\bibnamefont{Bergmann}},
  \bibinfo{journal}{Phys. Rev. B} \textbf{\bibinfo{volume}{77}},
  \bibinfo{pages}{104401} (\bibinfo{year}{2008}{\natexlab{a}}).

\bibitem[{\citenamefont{Bergmann}(2008{\natexlab{b}})}]{berg3}
\bibinfo{author}{\bibfnamefont{G.}~\bibnamefont{Bergmann}},
  \bibinfo{journal}{cond-mat} \textbf{\bibinfo{volume}{08050624}}
  (\bibinfo{year}{2008}{\natexlab{b}}).

\bibitem[{\citenamefont{Affleck et~al.}(2008)\citenamefont{Affleck, Borda, and
  Saleur}}]{affle2}
\bibinfo{author}{\bibfnamefont{I.}~\bibnamefont{Affleck}},
  \bibinfo{author}{\bibfnamefont{L.}~\bibnamefont{Borda}}, \bibnamefont{and}
  \bibinfo{author}{\bibfnamefont{H.}~\bibnamefont{Saleur}},
  \bibinfo{journal}{cond-mat} \textbf{\bibinfo{volume}{08020280}}
  (\bibinfo{year}{2008}).

\bibitem[{\citenamefont{Luo et~al.}(2005)\citenamefont{Luo, Verdozzi, and
  Kioussis}}]{doni2}
\bibinfo{author}{\bibfnamefont{Y.}~\bibnamefont{Luo}},
  \bibinfo{author}{\bibfnamefont{C.}~\bibnamefont{Verdozzi}}, \bibnamefont{and}
  \bibinfo{author}{\bibfnamefont{N.}~\bibnamefont{Kioussis}},
  \bibinfo{journal}{Phys. Rev. B} \textbf{\bibinfo{volume}{71}},
  \bibinfo{pages}{033304} (\bibinfo{year}{2005}).

\bibitem[{\citenamefont{Anderson}(2000)}]{pwa}
\bibinfo{author}{\bibfnamefont{P.~W.} \bibnamefont{Anderson}},
  \bibinfo{journal}{Physics Today} \textbf{\bibinfo{volume}{February}},
  \bibinfo{pages}{11} (\bibinfo{year}{2000}).

\bibitem[{\citenamefont{Goldhammer and Feenberg}(1956)}]{bw2}
\bibinfo{author}{\bibfnamefont{P.}~\bibnamefont{Goldhammer}} \bibnamefont{and}
  \bibinfo{author}{\bibfnamefont{E.}~\bibnamefont{Feenberg}},
  \bibinfo{journal}{Phys. Rev.} \textbf{\bibinfo{volume}{101}},
  \bibinfo{pages}{1233} (\bibinfo{year}{1956}).

\bibitem[{\citenamefont{Lippmann}(1956)}]{lipp}
\bibinfo{author}{\bibfnamefont{B.~A.} \bibnamefont{Lippmann}},
  \bibinfo{journal}{Phys. Rev.} \textbf{\bibinfo{volume}{103}},
  \bibinfo{pages}{1149} (\bibinfo{year}{1956}).

\bibitem[{\citenamefont{Bethe}(1999)}]{bethe}
\bibinfo{author}{\bibfnamefont{H.~A.} \bibnamefont{Bethe}},
  \bibinfo{journal}{Reviews of Modern Physics} \textbf{\bibinfo{volume}{71}},
  \bibinfo{pages}{S1} (\bibinfo{year}{1999}).

\end{thebibliography}
\end{document}